\documentclass[10pt, conference]{IEEEtran}
\usepackage{amsmath,amssymb,amsfonts}
\usepackage{graphicx}
\usepackage{pifont}
\usepackage{setspace}
\usepackage{enumerate}

\usepackage{amsthm}
\usepackage{xcolor}
\usepackage[figuresright]{rotating}

\usepackage{graphicx}
\graphicspath{{/}{fig/}}

\usepackage{array}
\usepackage{textcomp}
\usepackage{xcolor}
\usepackage{multirow}
\usepackage{booktabs}
\usepackage{siunitx}

\usepackage{mathtools}
\usepackage{breqn}
\usepackage{float}

\usepackage{pgfplots}
\pgfplotsset{compat=1.7}
\usepgfplotslibrary{groupplots}

\usepackage[style=base]{caption}
\usepackage{subcaption}

\usepackage{multirow,tabularx}
\usepackage{hyperref}
\usepackage{flushend}
\usepackage{algorithmic}
\usepackage[vlined, ruled, shortend]{algorithm2e}

\usepackage[capitalise]{cleveref}
\usepackage{svg}

\newbox{\bigpicturebox}

\newlength\figureheight
\newlength\figurewidth
\SetAlCapNameFnt{\footnotesize}
\SetAlCapFnt{\footnotesize}

\SetKwInput{kwInit}{Initilization}

\title{
    Distributed Ledger Technologies for Managing Heterogenous Computing Systems at the Edge \\
}

\author{
    \IEEEauthorblockN{
        \vspace{1em}
        Daniel Montero Hern\'andez\IEEEauthorrefmark{2},
        Jorge Pe\~na Queralta\IEEEauthorrefmark{2}
        and~Tomi Westerlund\IEEEauthorrefmark{2}%
    }
    \IEEEauthorblockA{
        \normalsize
        \IEEEauthorrefmark{2}\href{https://tiers.utu.fi}{Turku Intelligent Embedded and Robotic Systems (TIERS) Lab, University of Turku, Finland}.\\
        Emails: \textsuperscript{1}\{daanmh, jopequ, tovewe\}@utu.fi\\[+6pt]
    }
}

\begin{document}

\maketitle
\thispagestyle{empty}
\pagestyle{empty}

%%%%%%%%%%%%%%%%%%%%%%%%%%%%%%%%%%%%%%%%%%%%%%
%%                                          %%
%%           ABSTRACT AND TITLE             %%
%%                                          %%
%%%%%%%%%%%%%%%%%%%%%%%%%%%%%%%%%%%%%%%%%%%%%%

%%%%%%%%%%%%%%%%%%%%%%%%%%%%%%%%%%%%%%%%%%%%%%
%%                                          %%
%%                ABSTRACT                  %%
%%                                          %%
%%%%%%%%%%%%%%%%%%%%%%%%%%%%%%%%%%%%%%%%%%%%%%

\begin{abstract}%
    \label{sec:abstract}%
    The increased use of Internet of Things (IoT) devices —from basic sensors to robust embedded computers— has boosted the demand for information processing and storing solutions closer to these devices. Edge computing has been established as a standard architecture for developing IoT solutions, since it can optimize the workload and capacity of systems that depend on cloud services by deploying necessary computing power close to where the information is being produced and consumed. However, as the network scale in size, reaching consensus becomes an increasingly challenging task. Distributed ledger technologies (DLTs), which can be described as a network of distributed databases that incorporate cryptography, can be leveraged to achieve consensus among participants. In recent years DLTs have gained traction due to the popularity of blockchains, the most-well known type of implementation. The reliability and trust that can be achieved through transparent and traceable transactions are other key concepts that bring IoT and DLT together. We present the design, development and conducted experiments of a proof-of-concept system that uses DLT smart contracts for efficiently selecting edge nodes for offloading computational tasks. In particular, we integrate network performance indicators in smart contracts with a Hyperledger Blockchain to optimize the offloading on computation under dynamic connectivity solutions. The proposed method can be applied to networks with varied topologies and different means of connectivity. Our results show the applicability of blockchain smart contracts to a variety of industrial use cases.
\end{abstract}

\IEEEpeerreviewmaketitle

%%%%%%%%%%%%%%%%%%%%%%%%%%%%%%%%%%%%%%%%%%%%%%
%%                                          %%
%%                SECTIONS                  %%
%%                                          %%
%%%%%%%%%%%%%%%%%%%%%%%%%%%%%%%%%%%%%%%%%%%%%%
%%%%%%%%%%%%%%%%%%%%%%%%%%%%%%%%%%%%%%%%%%%%%%
%%                                          %%
%%              INTRODUCTION                %%
%%                                          %%
%%%%%%%%%%%%%%%%%%%%%%%%%%%%%%%%%%%%%%%%%%%%%%

\section{Introduction}\label{sec:introduction}

Edge computing is a paradigm of distributed computing and a well-established topological concept. Its primary objective is to optimize workload and capability of systems, by placing information processing capabilities closer to where things and people produce and consume said information~\cite{hamilton_edge_2018}.

\begin{figure}
    \centering
    \includegraphics[width=.49\textwidth]{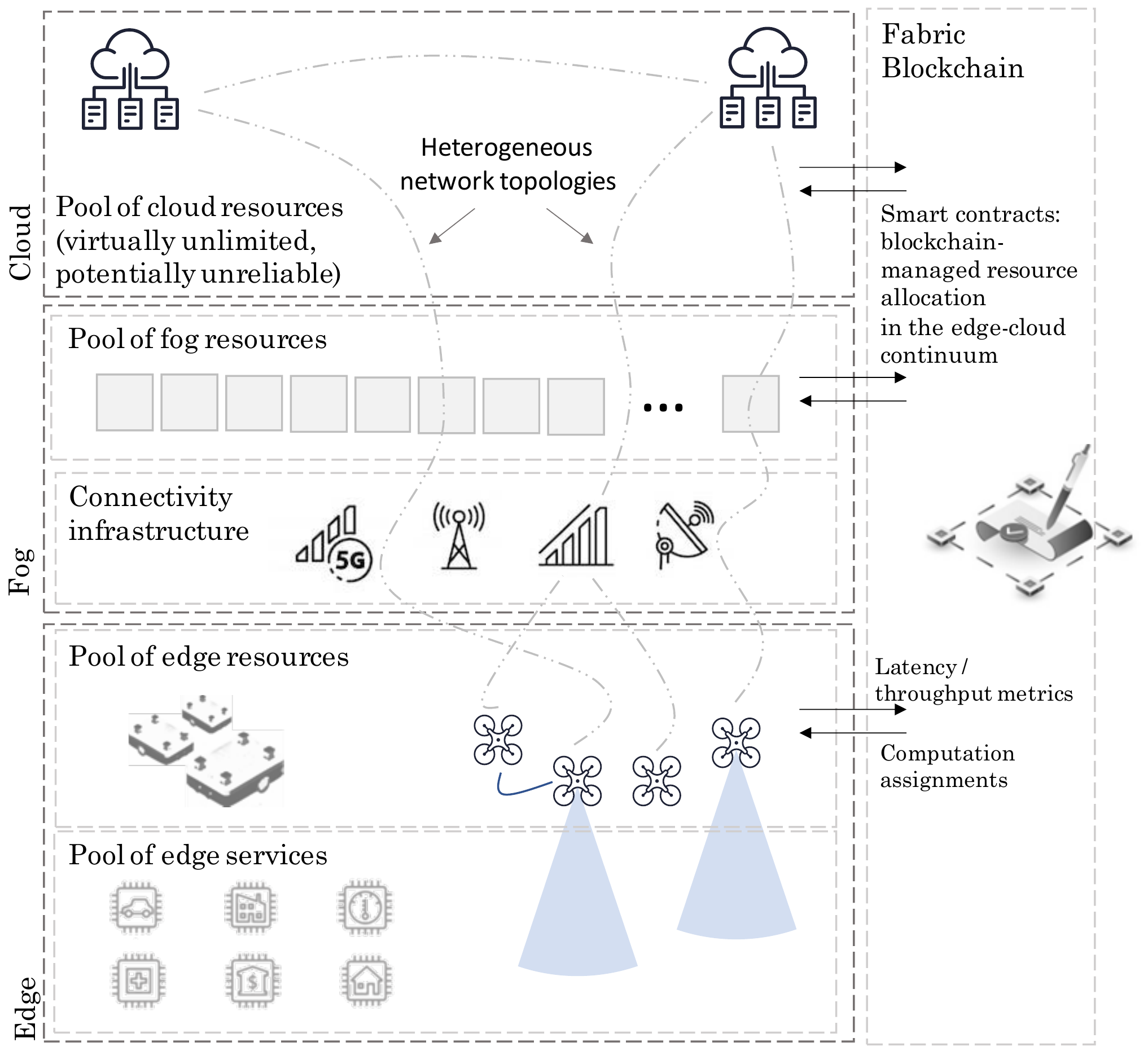}
    \caption{Conceptual illustration of the architecture and methods presented in this paper, where permissioned blockchain smart contracts aid in managing resources within the edge-cloud continuum.}
    \label{fig:concept}
\end{figure}

The crescent popularization of edge computing within IoT has been a topic of research in the past years~\cite{shi_edge}. This exponential growth of IoT has been leveraged by the continued interconnection of devices, computing resources and robots. An IoT network is made up of devices with network capabilities, which monitor data or bring intelligence to multiple domains~\cite{armDOS}. We are particularly interested in this research in Industrial IoT (IIoT) systems comprising mobile robots and other connected infrastructure within the scope of the RoboMesh project~\cite{westerlundintelligent}. Such a hybrid system often involves various network topologies and connectivity technologies, meaning that some nodes are connected locally while other connections rely on cloud servers~\cite{queralta2020end}. This type of system is illustrated in Fig.~\ref{fig:concept}. 

Over the past decade, multiple research efforts have been put towards resource allocation in edge computing systems~\cite{mao2017mobile}. In most cases, this has been focused around managing resources in local networks or at the radio access network (RAN) layer~\cite{mach2017mobile}. Within the technologies that have been proposed to manage computational resources across edge and cloud, and within the RAN infrastructure in mobile edge computing~\cite{queralta2020enhancing}, is blockchain technology~\cite{blockchain_edge_integration, queralta2020blockchain, queralta2021blockchain}. In this paper, we look into the integration of network provenance indicators for optimizing the distribution of a computational load with various connectivity solutions and hybrid network topologies. To do so, we rely on a next-generation industrial-focused blockchain solution: Hyperledger Fabric. Fabric blockchain networks, within the wider domain of distributed ledger technologies (DLTs) provide a scalable, dependent and managed solution that meets the performance and security standards of industrial applications and use cases~\cite{lu2017industry}.

From the conceptual illustration in Fig.~\ref{fig:concept}, we specifically look at managing resources across edge, fog and cloud using a Hyperledger Fabric Blockchain in hybrid systems with heterogeneous network topologies. A daemon deployed across all devices and connected to the blockchain routinely measures network performance indicators (e.g., latency and throughput between all pairs of nodes) to optimize the resource allocation mechanisms.

In summary, as a means of bringing together some of the latest key technical developments, this manuscript delves into the design and development of a Hyperledger Fabric blockchain system that through smart contracts (SC) will: collect resource data from edge nodes; measure network latency between nodes and devices in the network; and handle the data storage and edge node selection for offloading. The main contributions of this work are:

\begin{enumerate}[i.]
    \item the design and implementation of smart contracts for managing an inventory of devices; gathering device data; storing latency measurements; selecting the best edge node for computational offloading; and 
    \item the deployment and experimentation of the proposed solution with different connectivity solutions for edge nodes.
\end{enumerate}

%%%%%%%%%%%%%%%%%%%%%%%%%%%%%%%%%%%%%%%%%%%%%%
%%                                          %%
%%              RELATED WORKS               %%
%%                                          %%
%%%%%%%%%%%%%%%%%%%%%%%%%%%%%%%%%%%%%%%%%%%%%%

%\newpage
\section{Related Work} \label{sec:related_work}

Both Edge Computing and Distributed Ledger Technologies are currently relevant and novel topics in both industry and academia. 

Multiple works in the literature have reviewed the potential for distributed systems in the IoT brought by the convergence of blockchain technology and the edge computing paradigm~\cite{yang2019integrated, danzi2019delay, khezr2019blockchain, nguyen2020blockchain}. Other works that focus on either of the fields also mention the potential of integrating these technologies to bring more secure, dependable, and decentralized distributed computing systems to reality~\cite{gao2018survey, joshi2018survey, khan2019edge, moura2020fog}. In many of these works, a key issue identified for real-world deployment of blockchain technology is scalability, an inherent problem to traditional blockchain solutions that is however being solved in next-generation DLT frameworks~\cite{yang2019integrated, salimi2022secure, salimi2022towards, torrico2022uwb, keramat2022byzantine}. It is also worth noting that a substantial portion of the literature refers to solutions built around the Ethereum blockchain, which until recently has relied in proof-of-work (PoW) leading to computational and scalability limits in its root~\cite{danzi2019delay, ethereum2017goals}. This is, however, changing with both new Ethereum consensus algorithms and other DLT solutions~\cite{ethereum2018sharding, ethereum20specification}.

The management of Edge systems is a challenging topic and research regarding the use of DLT as an assisting technology is growing in popularity. Security, resiliency, and permissioned multi-actor participation; characteristics of DLT~\cite{itu_dltfg_2019, queralta2022secure}, can help overcome the challenge of consensus in the collaborative networks. The system must be able to reach an agreement regarding the state of the network and the availability of the resources. Consensus is one of the biggest challenges that we are faced with when planning and designing systems that can be large in scale and must also communicate with potentially heterogeneous computing resources while attempting to bring results in a low-latency basis. For the remainder of this section, we explore and review the different architecture proposals and prototypes of DLT frameworks for Edge Computing and Sensing systems.

\subsection{DLT and Edge Computing}

As the advancement of computing technologies allows for extension of systems, Edge Computing, and other similar implementations, such as Fog Computing, become more prevalent as ways to expand the use of the Cloud. Similar to the Cloud, Edge Computing has the objective of assisting the user by providing computation power, data storage and application services in a manner that maintains lower latency and improves the perceived quality of service. While there exist many benefits to the technology, the literature points out the challenges that such implementations have regarding the security and privacy of the information that is processed, due to the constant migration of services across edge nodes ~\cite{blockchain_edge_integration}. 

Literature that explores the application of DLT-backed systems is becoming more prevalent as the technology becomes more frequent due to the numerous applications that make use of it, ranging from smart grids to localized communication of IoT devices. One of the key concepts that is used to tie blockchain technology to the management of Edge Nodes is the guarantee of data integrity and validity, however, it is also mentioned that the cryptographic workload required can be computer-intensive depending on the consensus algorithm that is used ~\cite{blockchain_meets_edge}.

\subsection{DLT and Robotics Systems}

Recent works have showcased the potential of Hyperledger Fabric in robotic systems. Salimi et al. implemented a proof-of-concept DLT framework, and the authors conclude that the integration of robotics applications with Hyperledger Fabric can have minimal impact on the utilization of computational resources~\cite{salimi2022towards}. The same framework has also been deployed for multi-robot collaboration~\cite{salimi2022secure} and multi-robot role allocation applications~\cite{torrico2022uwb}. Earlier works have shown the potential of Ethereum smart contracts in swarm robotic use-cases, mainly in the detection of Byzantine agents~\cite{ferrer2018blockchain, strobel2018managing}.

In some of our previous works, we have also presented an architecture that enhances the autonomous operations of connected robots and vehicles with blockchain as the enabling technology to provide services and manage resources in a transparent and secure way ~\cite{queralta2020enhancing}, as well as the higher-level role that blockchain can play in robot swarms~\cite{queralta2020end, queralta2020blockchain}.

Another family of DLTs relevant to robot systems is directed acyclic graph (DAG)-based technologies, owing to the wider flexibility in terms of network topologies. IOTA is one of the most mature DLTs in this domain. The recent introduction of smart contracts within chains that are anchored to partitions of the DAG enables the design and implementation of distributed byzantine-tolerant decision-making processes, as in Ethereum, but that are also tolerant to network partitions under certain conditions. Recent works show that there is potential at the integration of ROS\,2 and IOTA, with the first byzantine-tolerant and partition-tolerant DLT-based solution for distributed consensus in multi-robot systems~\cite{keramat2022byzantine}.
%%%%%%%%%%%%%%%%%%%%%%%%%%%%%%%%%%%%%%%%%%%%%%
%%                                          %%
%%        PROBLEM DEFINITION                %%
%%                                          %%
%%%%%%%%%%%%%%%%%%%%%%%%%%%%%%%%%%%%%%%%%%%%%%

%\newpage
\section{Background}

The main objective of this manuscript is to describe the development of the Edge Management and Offloading System by exploiting the characteristics of different distributed computing paradigms, hence the used technologies will be explained in this section. Furthermore, the software and hardware used in the deployment of the system will be presented and distinguished.

\subsection{Distributed Computing and Edge Computing}

Through the years, multiple definitions for Distributed Computing have appeared in literature and rarely do these definitions agree with one another. The differences between these definitions usually lies in the scope of the distributed computing system that is being described by the authors. For the purpose of this work, we have settled on a combination of the definitions used by van Steen \cite{steen_tanenbaum_2017} and Coulouris \cite{coulouris_2012}: \endquote{"A distributed system is a collection of autonomous networked computing elements, that appears to its users as a single coherent system, that communicate and coordinate their actions by passing messages"}.

Edge Computing is a distributed computing topology and design paradigm and is commonly used to design the architecture of applications. This particular network topology supports the introduction of computing applications and services as close as possible to the source of the data that will be computed. This source of data depends on the particular use case of the system but can generally be referred to as the end-user of a system; and accordingly, this section of the network where the data sources and end-users are located is referred to as The Edge.~\cite{hamilton_edge_2018}.

Typically, Edge Computing serves as an additional layer between end-devices and Cloud services. The main reason why the architecture of a service will include Edge capabilities is to diminish the latency that exists between the two aforementioned layers. However, the benefits of this paradigm depend heavily on the application.

\subsection{DLT and Blockchain}

Distributed ledgers are consensually shared and synchronized databases. In contrast to a centralized ledger, distributed ledgers are less prone to cyber-attacks and fraud, since they do not suffer from having a single point of failure~\cite{qian2018towards, su2014decentralized, ayoade2018decentralized}. In a blockchain network the data stored in the ledger is updated in real-time via the consensus of the different nodes present in the network~\cite{swan2015blockchain, underwood2016blockchain}. As information is introduced by the users, the transactions are collected in groups, which are referred to as blocks. Once the system determines that the block has enough transactions, the block is closed and linked to the previous block, forming a chain of data, unlike a regular database table. The result of this configuration is an irreversible timeline of data transactions, which is why it is generally asserted that once data is part of the chain, it cannot be removed or edited.~\cite{hayes_blockchain_2022, cachin2016architecture}. There are other DLT implementations that differ from the Blockchain, but these are not relevant to the scope of this document.

Another important characteristic and differentiator  between DLTs is the consensus algorithm they use, and how the network and the permissions are configured. The most popular Blockchain implementations are known as permissionless, in which any device can join a network and participate in the activities and processes. This implementation allows for anonymous clients to participate, and consensus is typically reached via Proof of Work or Proof of Stake; Whereas in permissioned implementations the operator verifies and selects the entry of participants; Some types are also considered private, but for effects of this document, these do not count as fully distributed ledgers~\cite{seth_permissions_2022}. 

\subsection{Hyperledger Fabric}
Hyperledger Fabric is an open-source, modular and general-purpose DLT framework that offers identity management and access control features. As presented in the Hyperledger Foundation website\footnote{\url{https://hyperledger-fabric.readthedocs.io/en/latest/}}: "Hyperledger Fabric is an enterprise-grade permissioned distributed ledger framework for developing solutions and applications".

Additionally, it can operate a blockchain amongst a set of known, identified and often vetted participants that allows for secure interactions between entities that have a common goal, but do not necessarily fully trust each other. The identification of participants permits the use of consensus protocols that are less resource intensive. Hyperledger Fabric uses RAFT, a Crash Fault Tolerant (CFT) \footnote{\url{https://hyperledger-fabric.readthedocs.io/en/latest/orderer/ordering_service.html}} consensus algorithm, similar but more lightweight than the Practical Byzantine Fault Tolerant (PBFT) consensus algorithm.

In Fabric, smart contracts are written in Chaincode (CC), which is the software that defines the assets and all the instructions regarding the transformation of said assets and is installed on peers and then defined and used on one or more channels. The CC can be referred to as the business logic layer\footnote{\url{https://hyperledger-fabric.readthedocs.io/en/latest/glossary.html}}. These functions are initiated by a transaction proposal from a client, and result in a set of key-value writes which can be submitted in the network for ledger state propagation.

Other noteworthy properties of Hyperledger Fabric are that the execution environment is containerized in Docker, and the framework supports conventional high-level programming languages, like Java, JavaScript and Go, for the development of Smart Contracts and Applications.
%%%%%%%%%%%%%%%%%%%%%%%%%%%%%%%%%%%%%%%%%%%%%%
%%                                          %%
%%              METHODOLOGY                 %%
%%                                          %%
%%%%%%%%%%%%%%%%%%%%%%%%%%%%%%%%%%%%%%%%%%%%%%
\section{Methodology}

\begin{figure}
    \centering
    \includegraphics[width=.49\textwidth]{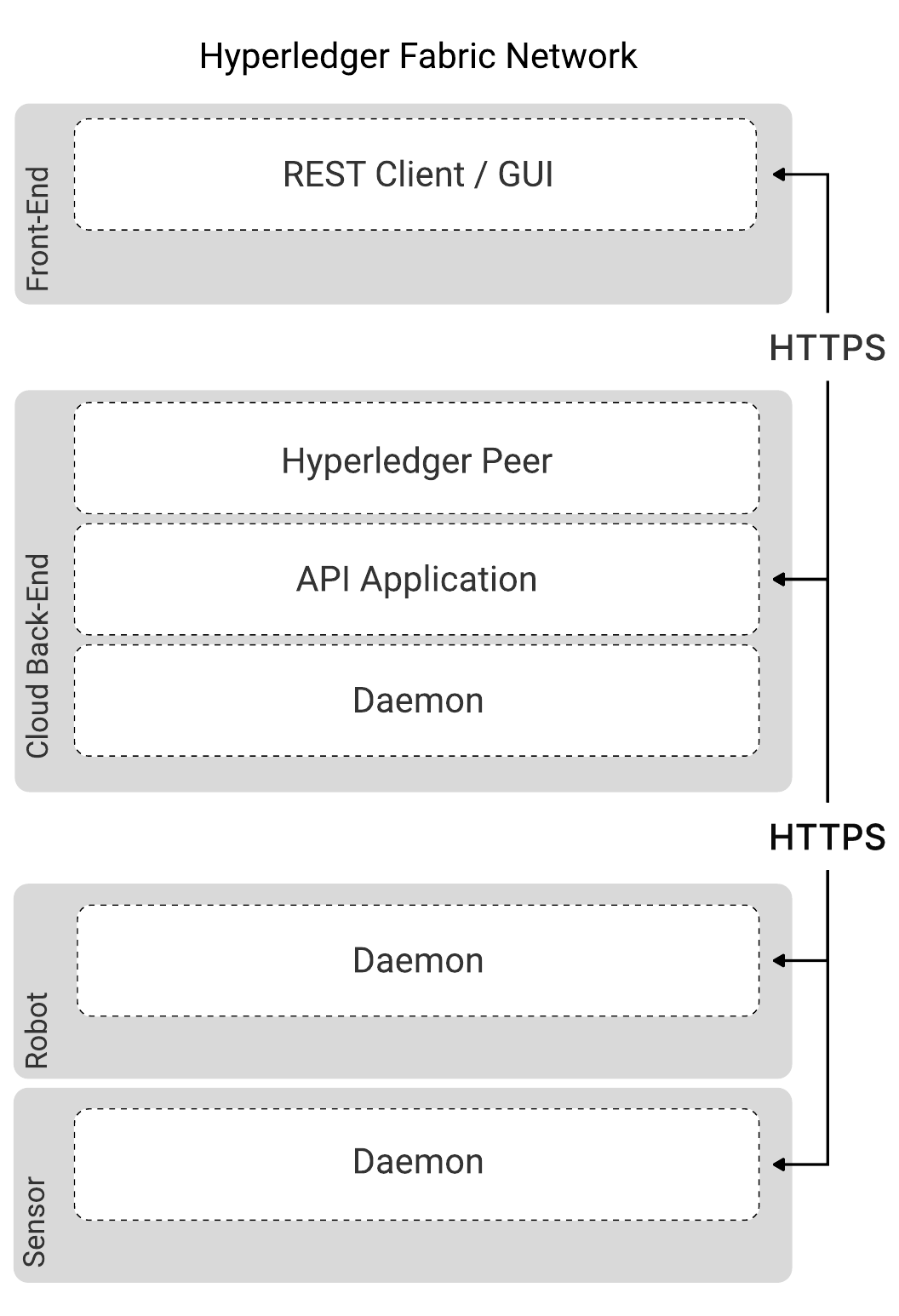}
    \caption{Diagram of the architecture.}
    \label{fig:architecture}
\end{figure}

In this section we detail the design of the solution along with the procedures for latency measurement and Edge Node selection. A diagram of the system is outlined in Fig.~\ref{fig:architecture}. 

The requirements that lead to the design of this solution are as follows:
\begin{itemize}
    \item \textbf{\textit{Collaboration}}: Resources are required to interconnect with various other devices and share information between each other.
    \item \textbf{\textit{Time Sensitive Analysis}}: The solution must be capable of quickly capturing and analysing the resource availability in the network.
    \item \textbf{\textit{Security, Trust \& Reliability}}: In order to achieve long-term functionality, the data in the system must be stored in a manner that prevents tampering, without affecting the ability to be shared, the stability or the speed of retrieval.
    \item \textbf{\textit{Extensibility \& Scalability}}: Due to fast-paced introduction of new devices and technology, the design must allow for flexibility in terms of configuration and capabilities. The addition of devices to the network must be seamless, whether the new devices are meant to be nodes capable of processing data, or more IoT devices for the sensing system.
\end{itemize}

The platform that we set out to design and implement has the objective of securely storing the information of the devices from a Sensing System and the Edge Computing Resources that will be available for offloading tasks. The data that will be stored is a combination of authentication information, resource status obtained by monitoring hardware and the latency that exists between the devices. All the chaincodes and Go Fabric applications are freely available in our GitHub repository~\footnote{\url{https://github.com/TIERS/fabric-edge-node-selector}}.

\subsection{Platform}

As described by Fig.~\ref{fig:architecture}, the components are the RoboMesh Platform, the devices that make up the Sensing System and the Edge Resources.

The first component is the RoboMesh Platform, this system is responsible for the management of the data that the devices in the network generate. This system is the combination of two components, a Hyperledger Fabric Network (HFN) with Smart Contracts and a REST Service combined with the Fabric SDK to expose an API that can be consumed by REST Clients or Web Applications.

The Sensing System is what we will refer to throughout the document as the IoT devices that capture data. The devices, or the computer that they are connected to, will deploy a Dockerized Daemon through which they will be monitored.

The Edge Resources are the hardware components in the network with the capabilities to perform tasks of higher complexity. Similarly, these computing devices will execute the Daemon in order to have their resources monitored, and gather which IoT devices against which they must measure latency. Additionally, the Daemon will serve as the tool in charge of executing the offloading tasks.

\subsection{Hyperledger Fabric Smart Contracts}
The HFN is made up of four different Smart Contracts that store data collected from the devices where the Daemon is installed, such as the Edge Nodes and the Sensing Devices. Three of these smart contracts are made up of CRUD operations and filtering functions.
\begin{itemize}
    \item \textbf{\textit{Inventory Management}}: responsible for securely storing the authentication procedure for accessing all the devices in the network. The stored object also contains some characteristics of devices required to properly filter and sort the required resources.
    \item \textbf{\textit{Resource Collection}}: responsible for securely storing the historic state of the computing devices and producing the necessary analysis of the data.
    \item \textbf{\textit{Latency Collection}}: responsible for securely storing the historical latency information between the devices that the Daemon generates.
    \item \textbf{\textit{Offload Selection}}: responsible for selecting which server will take part in the offloading of a task, based on the latency and resource status stored by the previously described contracts. The algorithm that analyzes and selects the correct node is described by Algorithm \ref{alg:offloading}.
\end{itemize}

\begin{algorithm}[t]
    \small
	\caption{Offload Server Selection}
	\label{alg:offloading}
	\KwIn{\\
	    \hspace{1em}Device to Offload task from: $target$ \\
	    \hspace{1em}Task Properties: $taskProperties$ \\
            \hspace{1em}Minutes for analysis time frame: $minutes$
	}
	\KwOut{\\
	    \hspace{1em}Selected Server : $selected$; \\
            \hspace{1em}List of other Servers : $selectedServerList$; \\
	}
	\BlankLine
    $latencyAnalysis$ = $latencySC$.AnalyseLatencyToTarget($target$, $minutes$)\;
    \If {$taskProperties$.GPU == TRUE} {
        $serverList$ = $inventorySC$GetServerListGPU()\;
    }
    \Else {
        $serverList$ = $inventorySC$GetServerList()\;
    }
    $filteredServerList$ = removeUnusedServers($serverList$)\;
    $resourceAnalysis$ = []\;
    \tcp{\scriptsize{Analyse each server in list concurrently}}
    \For{server in filteredServerList}{
        $resource$ = $resourceSC$.AnalyseResources($server$)\;
        $resourceAnalysis$.append($resource$)\;
    }
    $selectedServerList$ = combineAnalysis($resourceAnalysis$, $latencyAnalysis$)\;
    $selectedServerList$.sortByValue("latency", "averageCPU")\;
    $http$.Post(applicationUrl, $selectedServerList$[0])\;
    return $selectedServerList$
\end{algorithm}

The process that selects the correct server, described with pseudo-code in Algorithm \ref{alg:offloading}, starts by obtaining the analysis of Latency Measurements previously performed at the targeted device. The system then obtains the list of devices identified as servers, depending on whether the property of the task requires the Edge Node to have a GPU. To ensure that only available devices will be in the selection, the list is filtered to only include those Nodes which have performed the latency measurement in the specified time frame.
Using the filtered list of devices, we obtain the analysis of hardware resources of each Node concurrently. Latency and Resource analyses are merged into a list of Selection Objects. The list is then sorted in ascending order by the Latency Average. In the case that two or more devices share the same average, the list is sorted in ascending order by CPU Average, then by Memory Usage Average and finally by the amount of Containers being executed. This sorting method ensures that the server with the best connection or more available resources is selected.

\subsection{Daemon and Latency Measurement}

In order to correctly select the Edge Node with enough hardware resources for offloading the tasks, the Daemon collects the hardware status information on the device. This is the data that is used for analysis in the Resources SC of the HFN. For the network latency analysis between the host and the selected targets, instead of relying on Ping, we authenticate into the system and simulate data transfer to more accurately predict the time-to-reply of the systems, as described in Algorithm \ref{alg:latency_measurement}.

\begin{algorithm}[t]
    \small
	\caption{Latency Measurement}
	\label{alg:latency_measurement}
	\KwOut{\\
	    \hspace{1em}List of Latency Results : $measurementResults$; \\
	}  
	\BlankLine
        \tcp{\scriptsize{Get Targets from Application}}
        $targetList$ = $http$.Get(applicationUrl)\;
        $measurementResults$ = []\;
        \tcp{\scriptsize{Perform tasks concurrently}}
        \For{target in targetList}{
            $startTime$ = $time$.Now()\;
            $sshConnection$ = $ssh$.Dial(target.Authentication)\;
            $sshConnection$.executeCommand("echo " + startTime)\;
            $readTerminal$ = $sshConnection$.readBuffer()\;
            \tcp{Checks validity of connection and value of string}
            \If {(readTerminal == startTime) \&\& (sshConnection == TRUE)} {
                $sshConnection$.close()\;
                $elapsedTime$ = $time$.Since(startTime).Milliseconds()\;
            }
            \Else {
                $elapsedTime$ = -1\;
            }
            $measurementResults$.append({"target": target.ID, "latency": elapsedTime})\;
        }
        \tcp{\scriptsize{Send results to Application}}
        $http$.Post(applicationUrl, measurementResults)\;
        return $measurementResults$\;
\end{algorithm}

The Latency Measurement functionality obtains a list of devices to which the program needs to connect. In a concurrent fashion, the algorithm starts tracking the execution time as it performs SSH connections to the devices with the obtained authentication data, and prints information to check the validity of each connection. If the connection is valid, the connection is closed, and the execution time is stopped. Once all the concurrent tasks are finished, the list of measurements is sent to the Application via an HTTP Post.

\begin{figure}
    \centering
    \includegraphics[width=0.48\textwidth]{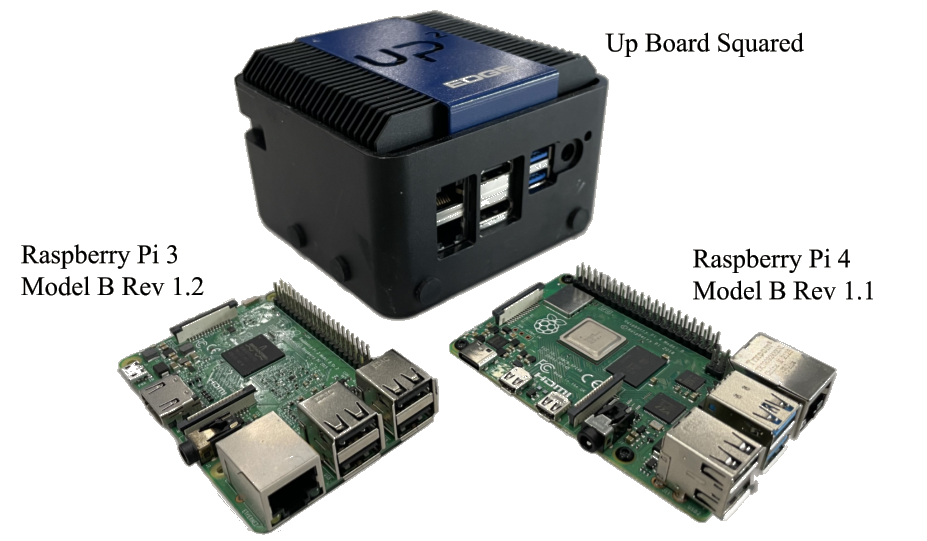}
    \caption{Hardware used in the experiments}
    \label{fig:hardware}
\end{figure}

\subsection{Hardware for Implementation and Experimentation}

In order to test the network with a variety of servers and devices to act as Edge Nodes or IoT devices, we have selected four different computers and one router, described below, and shown in Fig.~\ref{fig:hardware}.

\begin{itemize}
    \item \textbf{\textit{HFN Server}}: Host of the HF Test Network. Has an Intel i7-10750H (12) @ 5.000GHz CPU, 64GB of RAM and an NVIDIA GeForce RTX 2060 and Ethernet and WiFi capabilities.
    \item \textbf{\textit{Up Board Squared}}: Edge Node with an Intel Atom E3950 (4) @ 2.000GHz CPU and 8GB of RAM and Ethernet connection.
    \item \textbf{\textit{Raspberry Pi 3 Model B Rev 1.2}}: Edge Node with an ARMv7 BCM2835 (4) @ 1.200GHz CPU and 1GB of RAM and Ethernet connection.
    \item \textbf{\textit{Raspberry Pi 4 Model B Rev 1.1}}: Sensing Device with an ARMv7 BCM2711 (4) @ 1.500GHz CPU, 4GB of RAM and Ethernet and WiFi capabilities.
    \item \textbf{\textit{DLink DIR-882}}: Router with Ethernet and WiFi capabilities.
\end{itemize}

%%%%%%%%%%%%%%%%%%%%%%%%%%%%%%%%%%%%%%%%%%%%%%
%%                                          %%
%%              EXPERIMENTS                 %%
%%                                          %%
%%%%%%%%%%%%%%%%%%%%%%%%%%%%%%%%%%%%%%%%%%%%%%

\begin{figure}
    \centering
    \includegraphics[width=0.42\textwidth]{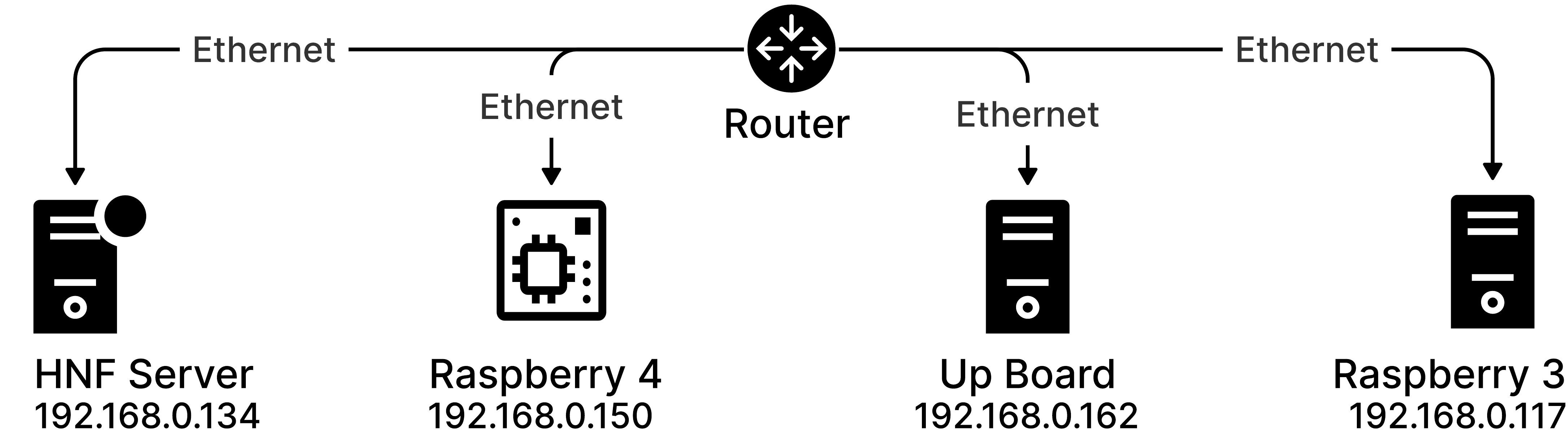} \\[+.5em]
    \footnotesize{Experimental setup \textit{(A)}.} \\[+1em]
    \includegraphics[width=0.42\textwidth]{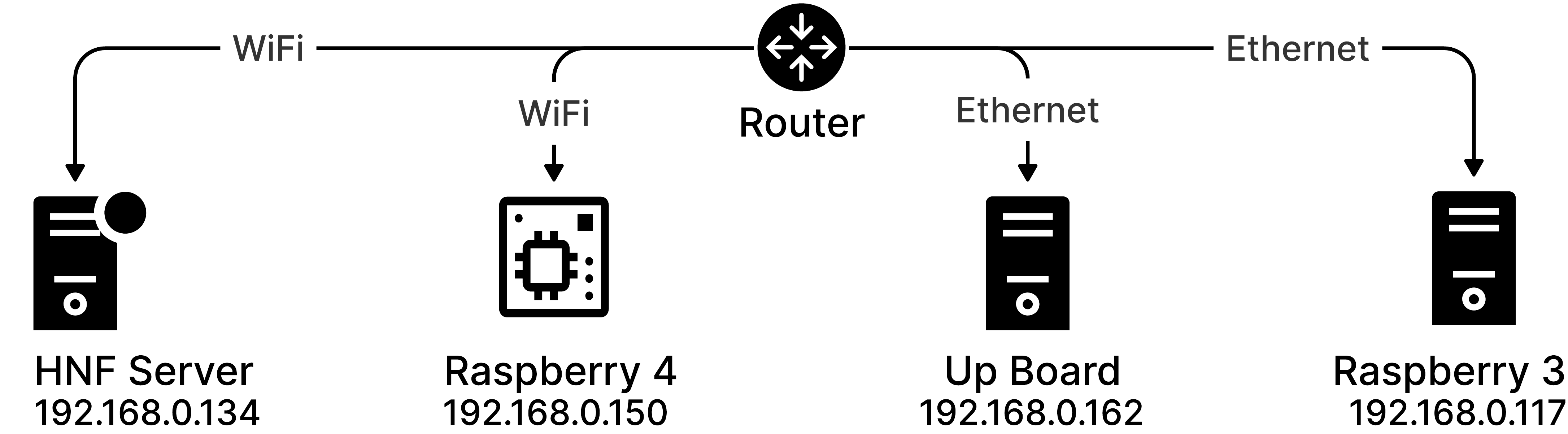}\\[+.5em]
    \footnotesize{Experimental setup \textit{(B)}.} \\[+1em]
    \includegraphics[width=0.42\textwidth]{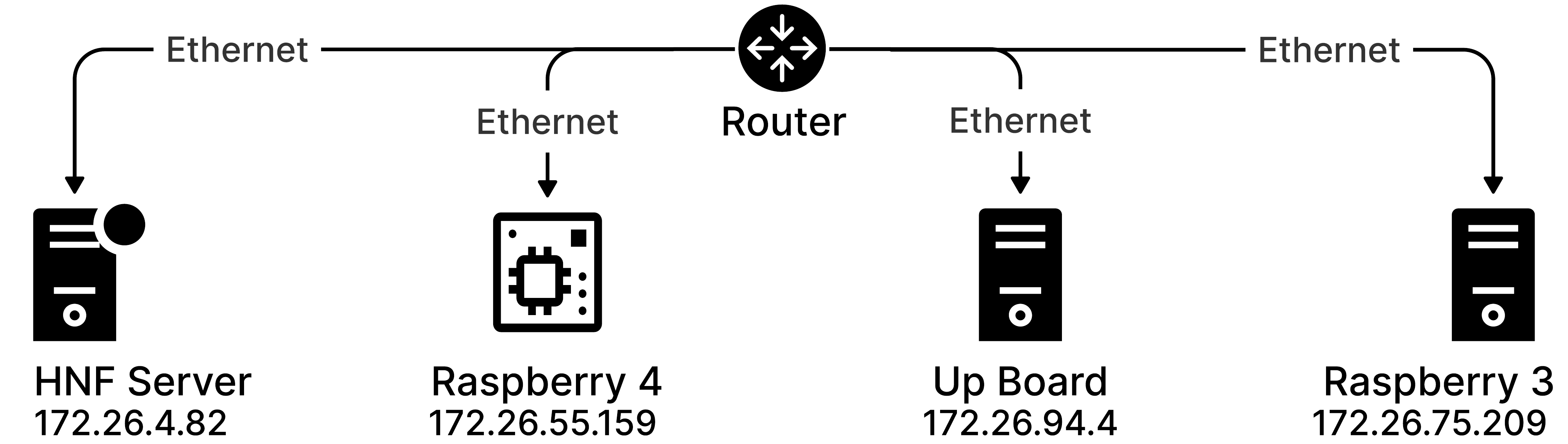}\\[+.5em]
    \footnotesize{Experimental setup \textit{(C)}.} \\[+1em]
    \includegraphics[width=0.42\textwidth]{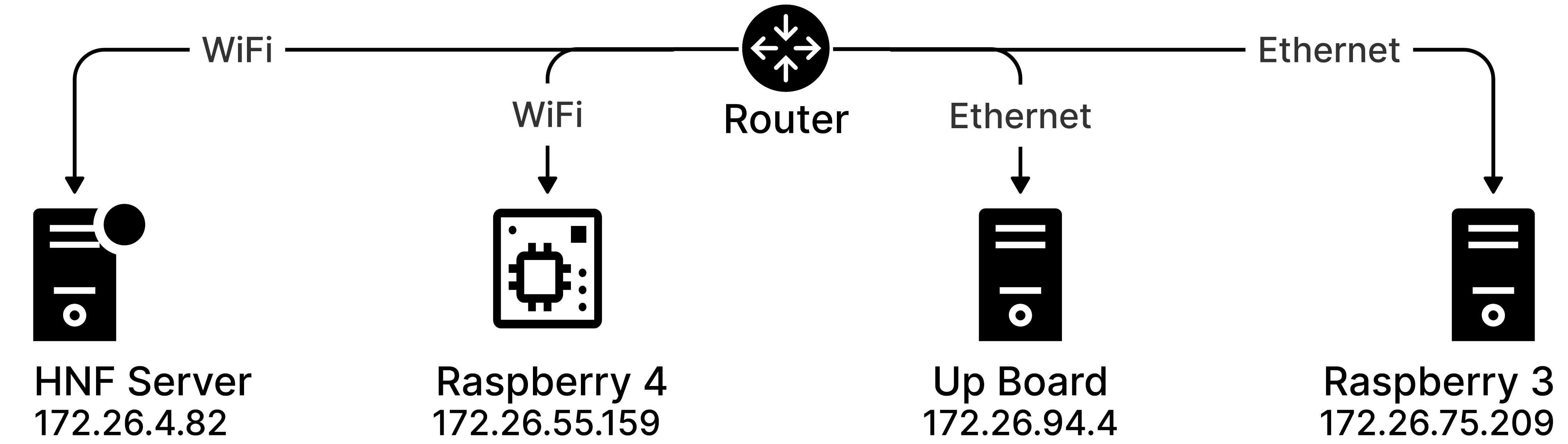}\\[+.5em]
    \footnotesize{Experimental setup \textit{(D)}.} \\[+1em]
    \includegraphics[width=0.42\textwidth]{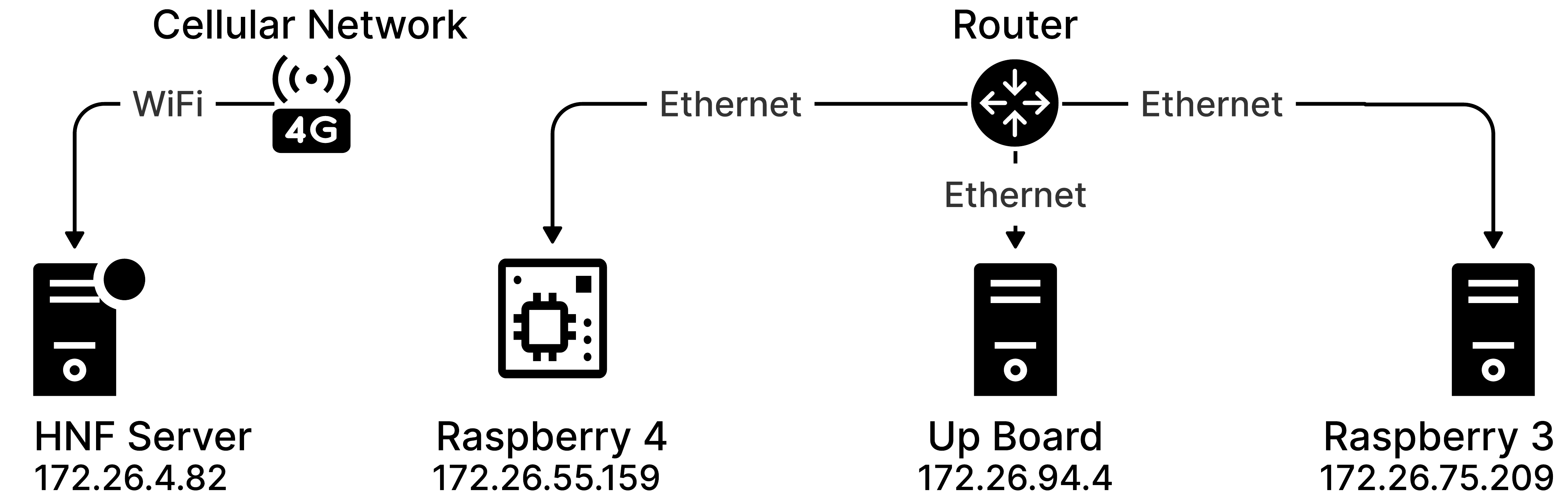}\\[+.5em]
    \footnotesize{Experimental setup \textit{(E)}.} \\[+1em]
    \caption{Illustration of the different experimental setups.}
    \label{fig:experimental_setups}
\end{figure}

\section{Experimental Results}

We have set up five scenarios, consisting of different network configurations, in order to demonstrate the platform's effectiveness in a real-world setting. Regardless of the scenario, every experiment performs an Edge Node Selection by executing the following steps:

\begin{enumerate}
    \item Configure devices, check connectivity and run the containerized Daemon.
    \item To prevent inaccurate analysis, the blockchain data and database are purged before restarting the HFN.
    \item Launch the Gateway Application.
    \item Since the Daemon is configured by default to execute operations every 30 seconds, we give the devices 15 minutes in order to gather enough data regarding the resource availability and latency measurements.
    \item The CPU-only Edge Node Selection feature is executed, and it performs a 10-min time frame analysis on the data.
\end{enumerate}

The data is retrieved and assessed after the selection has been made. The response is sent back to the user by the HTTP server of the Gateway application. In order to determine the effectiveness  and applicability of the proposed methods, we analyse the data created and stored by the Smart Contracts and the average duration of the HTTP operations. The rest of this section describes the different experimental setups, which are also illustrated in Fig.~\ref{fig:experimental_setups}.

\subsection{Local Network: Ethernet}

All devices are connected with an Ethernet cable to the router. All network operations are done using the Local Area Network IP Addresses assigned by the router's DHCP.

\begin{table}[H]
\centering
\small
\caption{\small{Local Network: Ethernet}}
\renewcommand{\arraystretch}{1.2}
    \begin{tabular}{@{} p{1.6cm}|p{1.5cm}|p{1.5cm}|p{1cm}|p{1.15cm}@{}}
    \toprule
    \textbf{Edge Node} & \textbf{Connection} &   \textbf{Latency}&   \textbf{CPU}&   \textbf{Memory}\\
    \midrule
    HFN Server   & Ethernet & 274.40\,ms & 4.97\% & 29.13\%\\
    Up Board   & Ethernet & 273.08\,ms & 2.08\% & 9.39\%\\
    Raspberry 3   & Ethernet & 280.80\,ms & 4.88\% & 14.36\%\\
    \bottomrule
    \end{tabular}
\label{table:exp1}
\end{table}

Under these conditions, the Up Board was selected as the preferred Edge Node. The data shows that the latency difference between nodes is negligible. In order to further contextualize, the \textit{ping} command indicates that the latency between both the selected server and the computer acting as the sensor was 0.659\,ms. The HTTP Server reported that the average duration for a read-only operation was 5.23\,ms, and the average of a write operation 1.67\,s.

\subsection{Local Network: Ethernet and WiFi}

Devices that are capable are connected to the router using WiFi in this scenario. All network operations are done using the Local Area Network IP Addresses assigned by the router's DHCP.

\begin{table}[H]
\centering
\small
\caption{\small{Local Network: Ethernet and WiFi}}
\renewcommand{\arraystretch}{1.2}
    \begin{tabular}{@{}p{1.6cm}|p{1.5cm}|p{1.5cm}|p{1cm}|p{1.15cm}@{}}
    \toprule
    \textbf{Edge Node} & \textbf{Connection} &   \textbf{Latency}&   \textbf{CPU}&   \textbf{Memory}\\
    \midrule
    HFN Server   & WiFi & 526.38\,ms & 5.70\% & 29.14\%\\
    Up Board   & Ethernet & 286.50\,ms & 2.05\% & 9.35\%\\
    Raspberry 3   & Ethernet & 335.30\,ms & 5.77\% & 13.88\%\\
    \bottomrule
    \end{tabular}
\label{table:exp2}
\end{table}

In this scenario, the preferred node was the Up Board, averaging 286\,ms of latency measurement. The \textit{ping} operation from the Up Board to the Raspberry 4 in this configuration averaged 2.22\,ms, a noticeable increase from the first experiment. The duration of HTTP operations remained close to those from the base experiment, averaging 5.58\,ms and 1.82\,s for read-only and write operations respectively.

\subsection{Local Network and VPN: Ethernet}

All devices were connected with an Ethernet cable to the router. All network operations are done using the IP Addresses assigned by the ZeroTier One VPN service.

\begin{table}[H]
\centering
\small
\caption{\small{Local Network and VPN: Ethernet}}
\renewcommand{\arraystretch}{1.2}
    \begin{tabular}{p{1.6cm}|p{1.5cm}|p{1.5cm}|p{1cm}|p{1.15cm} }
    \toprule
    \textbf{Edge Node} & \textbf{Connection} &   \textbf{Latency}&   \textbf{CPU}&   \textbf{Memory}\\
    \midrule
    HFN Server   & Ethernet & 276.18\,ms & 5.49\% & 29.62\%\\
    Up Board   & Ethernet & 306.62\,ms & 4.52\% & 9.05\%\\
    Raspberry 3 &  Ethernet & 357.07\,ms & 11.82\% & 13.57\%\\
    \bottomrule
    \end{tabular}
\label{table:exp3}
\end{table}

The usage of ZeroTier has impacted the latency in comparison to the first experiment, for the effects of our intended use of the system, we consider that the latency between devices is still acceptable. Under these conditions, the HFN Server was selected as the best Edge Node, and the \textit{ping} command averaged 1.16ms. Read-only and write operations seem to also have increased, albeit by a negligible amount, averaging 6.2\,ms and 2.11\,s respectively.

\subsection{Local Network and VPN: Ethernet and WiFi}

For this experiment, the devices with wireless capabilities are connected to the router using WiFi. All network operations are done using the IP Addresses assigned by the ZeroTier One VPN service.

\begin{table}[H]
\small
\centering
\caption{\small{Local Network and VPN: Ethernet and WiFi}}
\renewcommand{\arraystretch}{1.2}
    \begin{tabular}{@{}p{1.6cm}|p{1.5cm}|p{1.5cm}|p{1cm}|p{1.15cm}@{}}
    \toprule
    \textbf{Edge Node} & \textbf{Connection} &   \textbf{Latency}&   \textbf{CPU}&   \textbf{Memory}\\
    \midrule
    HFN Server   & WiFi & 331.05\,ms & 4.49\% & 29.84\%\\
    Up Board   & Ethernet & 341.86\,ms & 4.47\% & 9.03\% \\
    Raspberry 3   & Ethernet & 388.94\,ms & 12.39\% &13.79\%\\
    \bottomrule
    \end{tabular}
\label{table:exp4}
\end{table}

The selected Edge Node, in this case, was the HFN Server, with the latency measurement averaging at 331.05\,ms and the \textit{ping} command at 3.49ms. This experiment has made clear that using both WiFi and a VPN makes an impact, but the averages between the servers are consistent. The HTTP Server did not seem to have been too affected, with the durations averaging 6.02\,ms and 2.13\,s for read-only and write respectively.

\subsection{Mixed Network: Ethernet and WiFi}

In this experiment, the HFN Server is connected to a 4G cellular band. Other computers remain in the original network. All network operations are done using the IP Addresses assigned by the ZeroTier One VPN service.

\begin{table}[H]
\small
\caption{\small{Mixed Network: Ethernet and WiFi}}
\centering\renewcommand{\arraystretch}{1.2}
    \begin{tabular}{@{}p{1.6cm}|p{1.5cm}|p{1.5cm}|p{1cm}|p{1.15cm}@{}}
    \toprule
    \textbf{Edge Node} & \textbf{Connection} &   \textbf{Latency}&   \textbf{CPU}&   \textbf{Memory}\\
    \midrule
    HFN Server   & 4G & 4735.50\,ms & 1.98\% & 30.82\%\\
    Up Board   & Ethernet & 448.33\,ms & 2.50\% & 9.32\%\\
    Raspberry 3   & Ethernet & 445.21\,ms & 2.61\% & 12.64\%\\
    \bottomrule
    \end{tabular}
\label{table:exp5}
\end{table}

This final experiment confirms that even when using different networks, the system is still capable of finding an Edge Node. The HFN Server connected to a 4G Cellular Network has the most latency, and the Up Board connected to the same network as the Raspberry 4 was selected as the Edge Node. The \textit{ping} average between the selected node and the target device is 3.52\,ms. During this experiment, the averages for the read-only and write operations did change, with the average duration being 20.12\,ms and 3.11\,s respectively.

The increase of latency for the HTTP operations is directly related to the latency that now exists between the HFN Server and the rest of the devices.
%%%%%%%%%%%%%%%%%%%%%%%%%%%%%%%%%%%%%%%%%%%%%%
%%                                          %%
%%              CONCLUSION                  %%
%%                                          %%
%%%%%%%%%%%%%%%%%%%%%%%%%%%%%%%%%%%%%%%%%%%%%%

\section{Conclusion}\label{sec:conclusion}

We have presented the design and implementation of a Hyperledger Fabric Smart Contract based platform that can analyse the latency between devices and the resource availability of computing nodes in order to choose the most appropriate one to act as an offloading Edge Node. The experiments performed to test the efficacy of the system demonstrate that the Hyperledger Fabric permissioned network configuration enables the implementation of quick, secure, and dependable distributed applications in which participants do not necessarily completely trust one another.

The multiple experimental scenarios that we proposed and the data that we gathered from them indicate that through this system implementation we can achieve long-range capabilities, as the speed of the system remained manageable even with the use of commercially available 4G cellular networks and VPN services. It should be noted that the increased latency due to mixing networks can be addressed by deploying more instances of the Gateway Application in a production environment, where Fabric Peers are expected to also be deployed in the Edge Nodes.

An equally significant finding regarding the design of a Fabric Smart Contract based system is the structure of the digital asset that will be stored in the blockchain and database. Since the world state is recorded in a NoSQL engine, the possibilities of querying and joining data is limited. In order to achieve faster speeds when retrieving information, the assets must be designed in a way that they can later be indexed.

\subsection{Future Works}
We will expand the platform functionalities to determine whether Smart Contracts are also a feasible mechanism to share state information required for executing the offloading of tasks after having selected an Edge Node. The current system can accurately choose nodes based on their current state, and whether the task requires or not a GPU, but the system is not capable of starting the execution of the tasks, it does not have the capabilities to track the execution of said tasks nor does the Edge Node Selection happen autonomously. 

Currently, the functionalities of the system do not include a mechanism that ensures that the growth rate of the world state database will remain constant, which makes the long-term autonomy evaluation of the platform a challenge. We started designing methods in which older data can be archived in Cloud systems to avoid the indexing process of the ledger from deteriorating, thus maintaining the speed of the analysis.

Additionally, we would like to study the performance of the long-range capabilities of the system using a 5G Mesh Network, and the efficacy of the Offload Selection for robotics systems that require real-time processing.

%%%%%%%%%%%%%%%%%%%%%%%%%%%%%%%%%%%%%%%%%%%%%%
%%                                          %%
%%            ACKNOWLEDGMENT                %%
%%                                          %%
%%%%%%%%%%%%%%%%%%%%%%%%%%%%%%%%%%%%%%%%%%%%%%

\section*{Acknowledgment}

This research work is supported by the Academy of Finland's RoboMesh and AeroPolis projects (grant numbers 336061 and 348480).

%%%%%%%%%%%%%%%%%%%%%%%%%%%%%%%%%%%%%%%%%%%%%%
%%                                          %%
%%              BIBLIOGRAPHY                %%
%%                                          %%
%%%%%%%%%%%%%%%%%%%%%%%%%%%%%%%%%%%%%%%%%%%%%%
% \newpage
% \nocite{*}
\bibliographystyle{unsrt}
\bibliography{bibliography}

\end{document}